\documentstyle[12pt,epsf,aps]{revtex}
\begin{document}
\title{The beta decay of $^{44}$V}
\author{G. Mart\'{\i}nez-Pinedo and A. Poves}
\address{Departamento de F\'{\i}sica Te\'orica C-XI\\
         Universidad Aut\'onoma de Madrid\\
         E-28049 Madrid, Spain}
\date{February, 1993}
\maketitle
\begin{abstract}

   We have calculated the $\beta^+$ decay of the proton rich nuclei
$^{44}$V in a full fp-shell valence space.  We obtain a theoretical
half-life of 71~ms compared with the experimental value t$_{1/2}$ = 90
$\pm$ 25 ms.  Besides, we make predictions for the Gamow-Teller
strength functions corresponding to the decays of the 2$^+$ ground
state and 6$^+$ isomer state.

\end{abstract}

   The study of Gamow-Teller processes in light nuclei has received
considerable attention during the last years as a source of
information both on the nuclear structure and on the behaviour of the
axial vector weak current in the nuclear medium.  The systematic
analysis of (p,n) reactions, from which the Gamow-Teller (GT) strength
can be extracted, has led to the conclusion that the experimental
strength is depleted by a factor $\sim$ 0.5 relative to the Ikeda sum
rule \cite{gaar}.  This corresponds to a value of the ratio
$\frac{g_{A}}{g_{V}}$ close to 1, instead of the free value 1.251.
The same conclusion was reached in a careful analysis of the
individual beta decay rates in the sd shell \cite{brown}.
Nevertheless, the experiments carried out in the proton rich nuclei
$^{33}$Ar \cite{argon} and $^{37}$Ca \cite{calcio1} have questioned
these results.  Owing to their very large $Q_{\beta}$ values, these
decays explore a large fraction of the GT strength function, including
part of the GT giant resonance.  The experimental results for the
integrated strength are compatible with the theoretical predictions
without any renormalization.  Furthermore, in the $^{37}$Ca case the
GT strength functions obtained from the $\beta^{+}$ decay and from the
(p,n) reaction are different \cite{calcio2}.  These findings, even if
subject to debate \cite{good}, shed doubts on the adequacy of the
procedure of extraction of the GT strength from the (p,n) data.  The
conclusions of the sd shell analysis are also subject to criticism.
For, in most of the decays studied, the only available information
refers to low-lying levels.  Hence, what appears as a quenching of the
GT strength could be also interpreted as a systematic shift of the
calculated strength from the GT giant resonance towards the low-lying
levels.

   To clarify this situation it is important to study other
$\beta^{+}$ decays in proton rich nuclei with large Q$_\beta$ windows.
$^{44}$V is an excellent test case for several reasons.  The energy
released of its decay to $^{44}$Ti is very large (Q$_{\text{EC}}$=13.7
MeV); a theoretical description of A = 44 nuclei, using a reliable
effective interaction and a full $0\hbar \omega$ valence space, is
feasible; and not the least important, results from a scheduled
experiment at Ganil will come out soon \cite{ganil}.  This decay has
two peculiar features that are worth commenting:

\begin{itemize}

   \item[i)] $^{44}$V is the mirror of $^{44}$Sc, an experimentally
well know nuclei, therefore their wave functions are identical except
Coulomb effects.  When we compare the theoretical predictions for
$^{44}$Sc with the experimental data \cite{tablas}, we are checking
simultaneously the quality of the $^{44}$Sc and the $^{44}$V
wavefunctions.

   \item[ii)] $^{44}$Sc has an isomer state (6$^+$, t$_{1/2}$ =
2.44~d) at 0.271 MeV excitation energy.  Hence, $^{44}$V also will
have a 6$^+$ state at an excitation energy close to 270 keV.  Contrary
to what happens in the decay of the $^{44}$Sc isomer state, the
$^{44}$V isomer state is predicted to decay beta with a 100\%
branching ratio.

\end{itemize}

   We use the fp shell as valence space, without truncations.  The
effective interaction is a slightly modified version of the Kuo-Brown
\cite{kuo} interaction, denoted KB3 in \cite{poves}.  This interaction
gives a fairly good description of the spectroscopy of nuclei in the
lower and middle part of the fp-shell \cite{poves}.

\begin{figure}
\epsffile{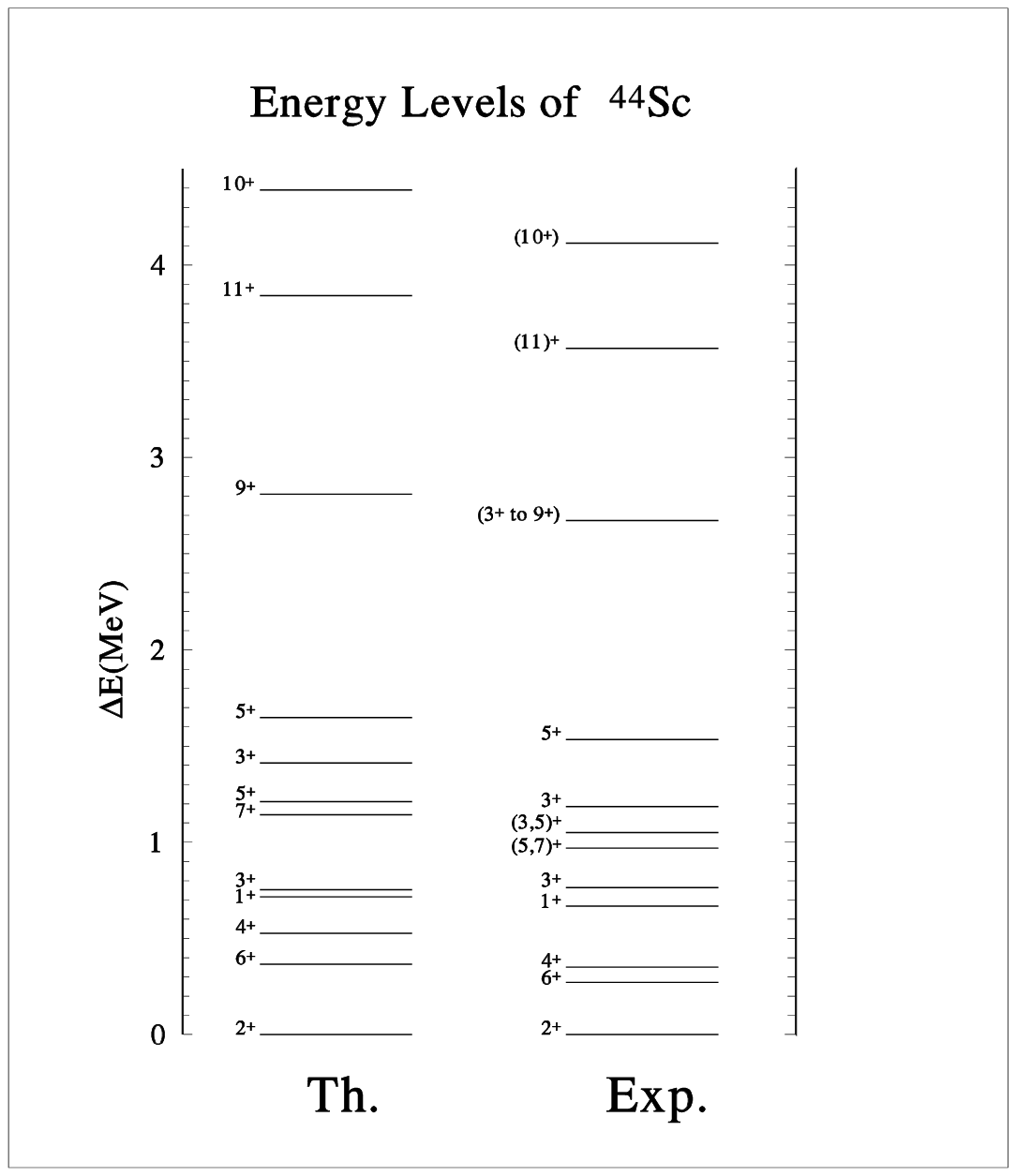}
\caption{Experimental and theoretical energy levels of $^{44}$Sc.}
\label{fig:en_sc}
\end{figure}

   To test the quality of the wave functions, we have calculated the
level scheme of $^{44}$Sc, the mirror of $^{44}$V.  In figure
\ref{fig:en_sc} we can see that the theoretical predictions compare
extremely well with the experimental results.  Next we examine the E4
transition from the 6$^+$ isomer of $^{44}$Sc to the ground state
(2$^+$).  Experimentally this decay mode proceeds with a 98.6\%
branching ratio, hence we disregard the contribution of the weak
branch to the half-life.  Using effective charges 1.5 for protons and
0.5 for neutrons the prediction for the half-life is t$_{1/2}$ = 2.96~d
in good agreement with the experimental result t$_{1/2}$ = 2.44~d.
We have also computed the electron capture branching ratio
(experimentally 1.4\%) obtaining a 1\% branching ratio to the only
available state, a 6$^+$ T = 2 at 3.285 MeV in $^{44}$Ca.  The only
state contributing significantly to the half-life of the $^{44}$Sc
ground state is the first excited state of $^{44}$Ca.  The predicted
half-life is 1.13~h while the experimental result is 3.9~h.

\begin{figure}
\epsffile{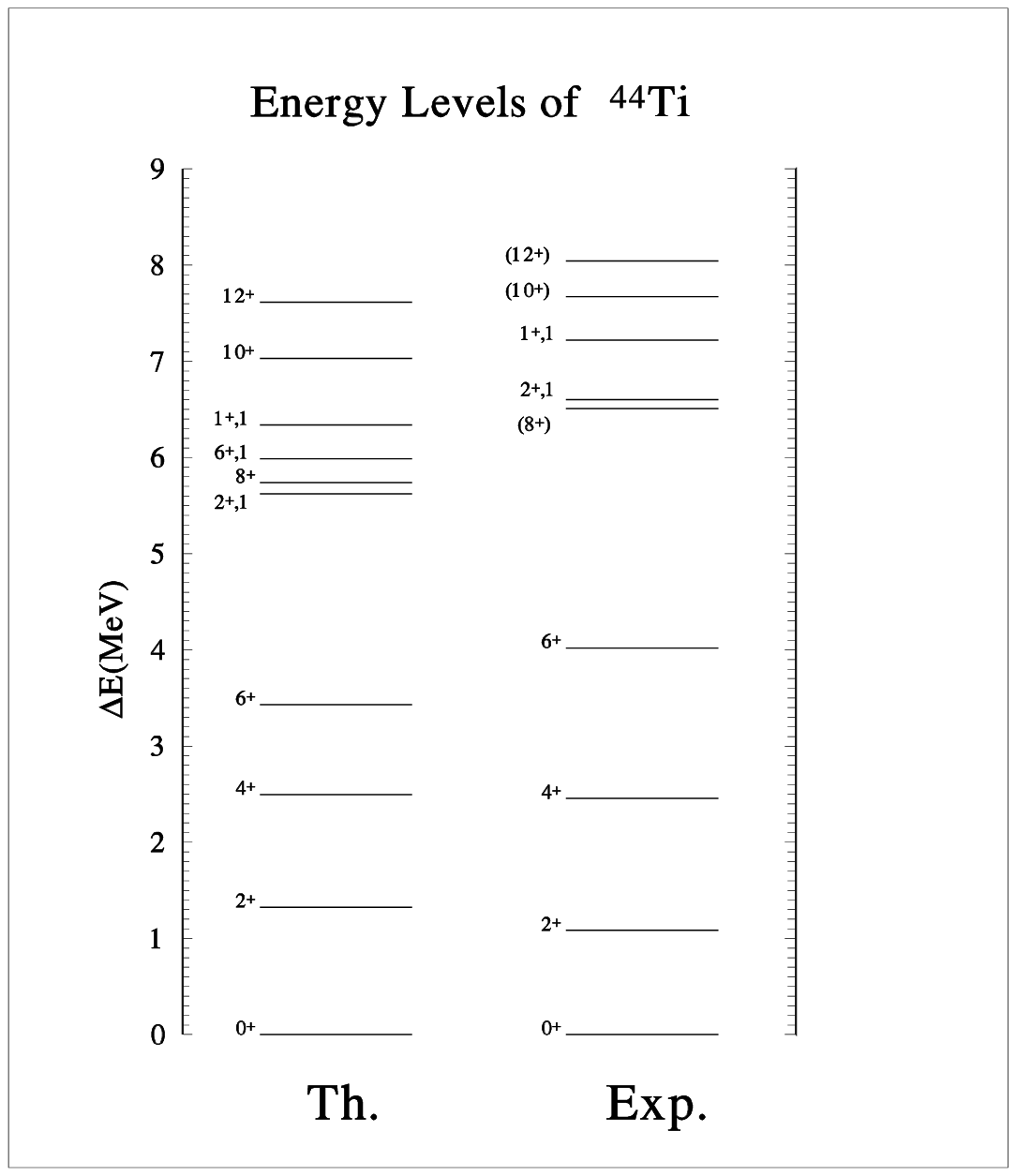}
\caption{Experimental and theoretical energy levels of
$^{44}$Ti.}
\label{fig:en_ti}
\end{figure}

   We come to the $^{44}$V $\longrightarrow$ $^{44}$Ti decay.  Our
description of the $^{44}$Ti spectroscopy is very satisfactory as it
can be gathered from figure \ref{fig:en_ti}.  The lowest $0^{+}$ and
$2^{+}$ states are underbound due to the absence of core excited
configurations in our approach.  This shift would have to be taken
into account when comparing the experimental and theoretical GT
strength functions.

\begin{figure}
\epsffile{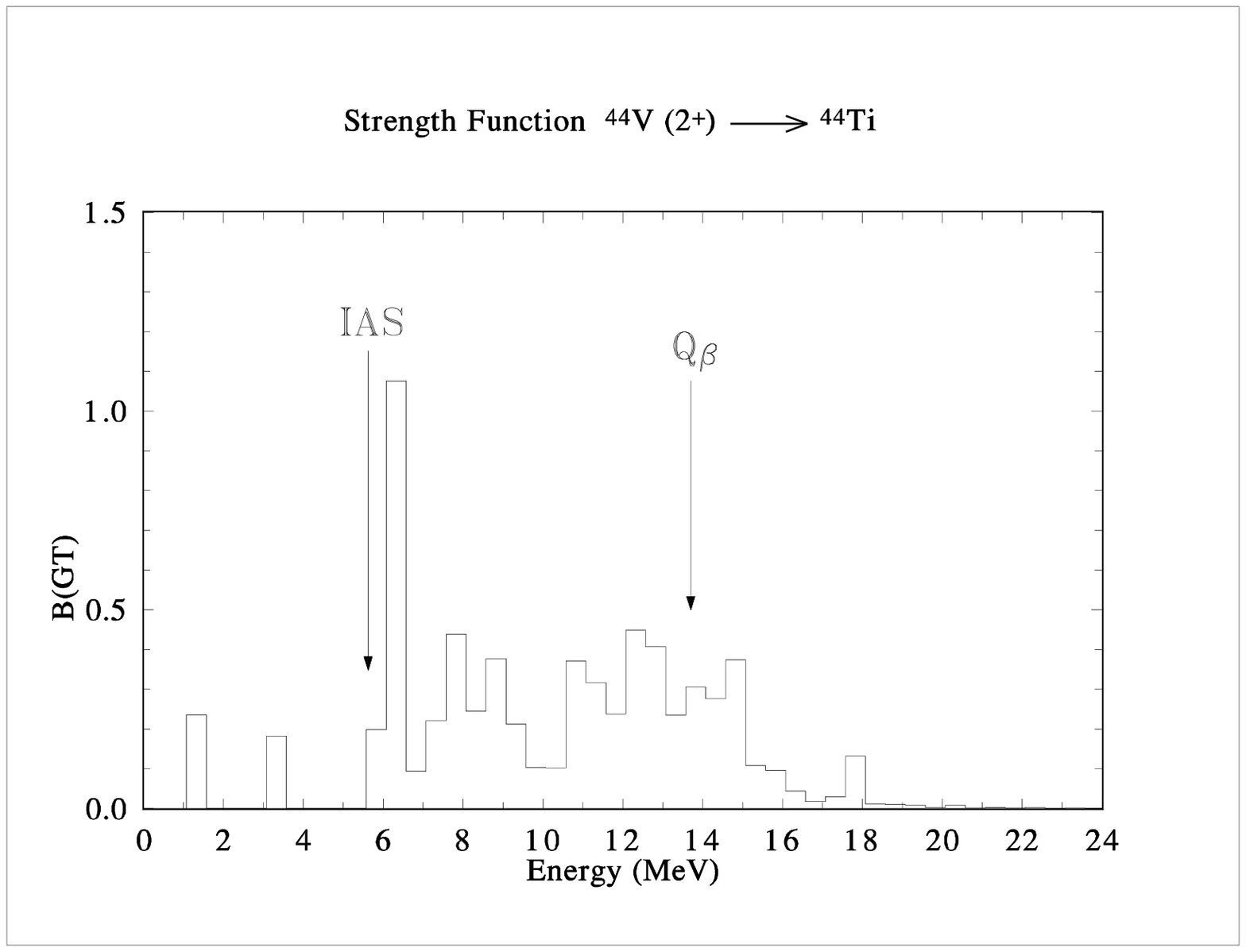}
\caption{$^{44}$V (2$^+$) $\stackrel{\beta^+}{\longrightarrow}$
$^{44}$Ti Gamow-Teller strength function.  B(GT) =
$\left(\frac{g_A}{g_V}\right)^2\ \langle \sigma \tau \rangle^2$.}
\label{fig:vj2ti}
\end{figure}

   The half-life of $^{44}$V is known experimentally to be t$_{1/2}$ =
90 $\pm$ 25 ms.  The calculation gives t$_{1/2}$ = 49 ms using the
bare value for $(g_A/g_V)$.  The agreement becomes excellent when we
use $\left(g_A/g_V\right)_{\text{eff}}$ = 0.77
$\left(g_A/g_V\right)_{\text{bare}}$, in this case the result is
t$_{1/2}$ = 71~ms.  Once again it appears that there is too much GT
strength in the low-lying states, the ones determining the half-life.
This result is in the line of the findings of ref. \cite{brown}.  The
predicted GT strength function is plotted in figure \ref{fig:vj2ti}.
Notice that almost the whole of the predicted strength (80\%) falls
into the $\beta^+$ energy window.

\begin{figure}
\epsffile{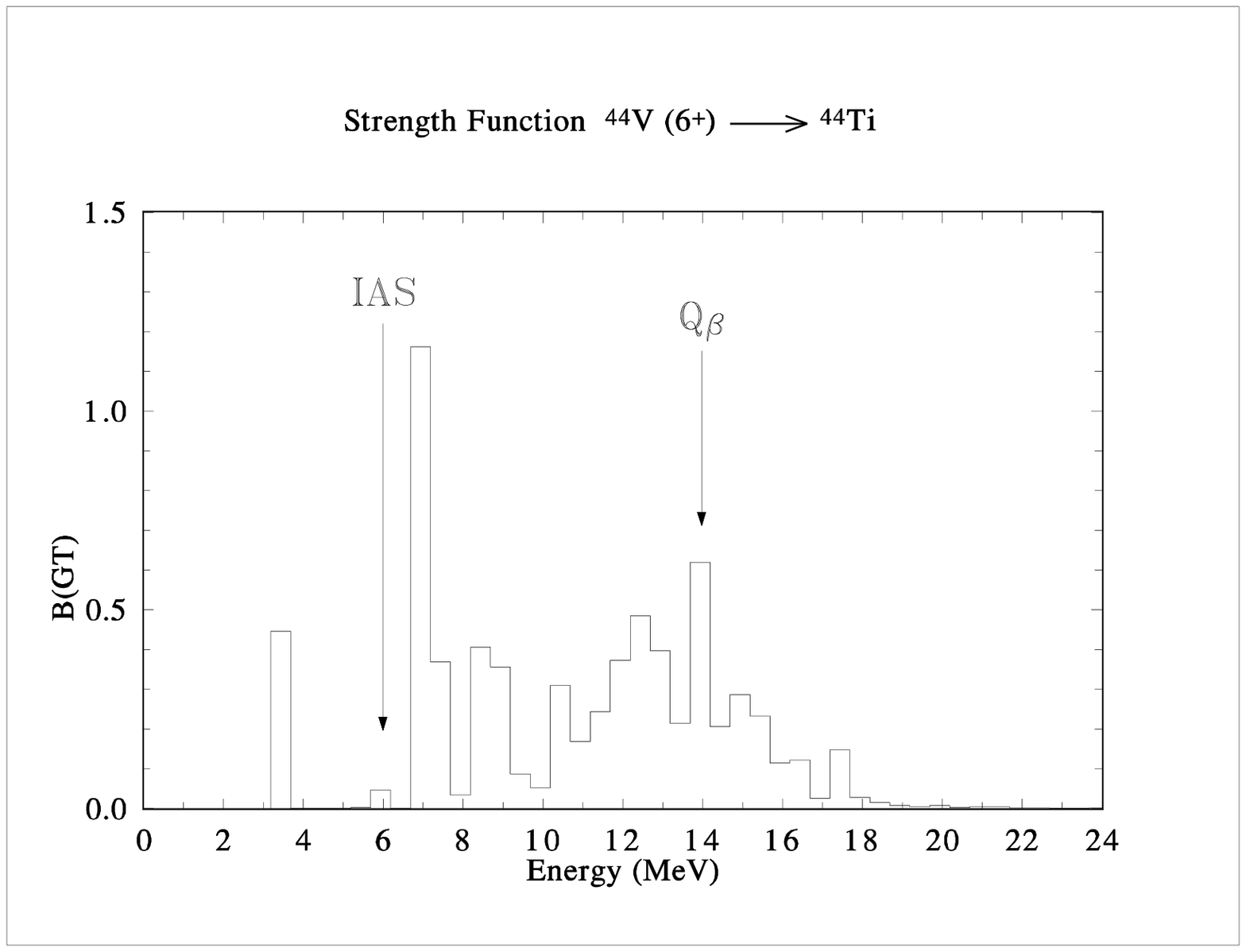}
\caption{$^{44}$V (6$^+$) $\stackrel{\beta^+}{\longrightarrow}$
$^{44}$Ti Gamow-Teller strength function.  B(GT) =
$\left(\frac{g_A}{g_V}\right)^2\ \langle \sigma \tau \rangle^2$.}
\label{fig:vj6ti}
\end{figure}

   Lets examine the behavior of the 6$^+$ isomer state of $^{44}$V.
The E4 transition to the ground state is predicted to proceed with a
half-life t$_{1/2}$ = 1.08 days, using effective charges 1.5 for
protons and 0.5 for neutrons.  Due to the much larger value of the
Q$_\beta$ and to the availability of T = 0 and T = 1 states to decay
to, the weak branch is dominant; t$_{1/2}$ = 66 ms using the bare
value for $(g_A/g_V)$.  The predicted Gamow-Teller strength function
for the 6$^+$ decay is plotted in figure \ref{fig:vj6ti}.  Most of the
strength can be explored by the $\beta^+$ process.  If the ground
state and the 6$^+$ isomer state are simultaneously fed in an
experiment we shall observe both decays in parallel with similar
half-lifes.

   In conclusion, we have computed the Gamow-Teller strength functions
of the ground state and of the isomer 6$^+$ state of $^{44}$V.  These
strength functions can be explored almost completely by the $\beta^+$
decay.  The comparison with the results of the experiment planned at
Ganil will help us to understand better the renormalization of the
axial vector weak current in nuclei.

\acknowledgements

   We thank G. Walter for discussions.  This work has been supported
partly by DGICYT (Spain) under contract PB89-164.

\end{document}